# Veloce Rosso: Australia's new precision radial velocity spectrograph

James Gilbert[a], Christoph Bergmann[b], Gabe Bloxham[a], Robert Boz[a], Robert Brookfield[c], Tom Carkic[a], Brad Carter[d], Scott Case[e], Vladimir Churilov[e], Michael Ellis[a], Gaston Gausachs[a], Luke Gers[a], Doug Gray[c], Nicholas Herrald[a], Michael Ireland[a], Damien Jones[f], Yevgen Kripak[e], Jon Lawrence[e], Ellie O'Brien[a], Ian Price[a], Matthew Robertson[a], Christian Schwab[g], Chris Tinney[b], Annino Vaccarella[a], Colin Vest[a], Duncan Wright[d], Ross Zhelem[e]

[a] Australian National University, Mt Stromlo Observatory, Cotter Road, Weston Creek, ACT 2611, Australia
[b] Exoplanetary Science at UNSW, School of Physics, UNSW Sydney, NSW 2052, Australia
[c] Anglo-Australian Telescope, Siding Spring Observatory, Coonabarabran, NSW 2357, Australia
[d] University of Southern Queensland, 37 Sinnathamby Boulevard, Springfield Central, QLD 4300, Australia
[e] Australian Astronomical Observatory, 105 Delhi Road, North Ryde, NSW 2113, Australia
[f] Prime Optics, 17 Crescent Road, Eumundi, QLD 4562, Australia
[g] Macquarie University, Balaclava Road, North Ryde, NSW 2109, Australia

## ABSTRACT

Veloce is an ultra-stable fibre-fed R4 echelle spectrograph for the 3.9 m Anglo-Australian Telescope. The first channel to be commissioned, Veloce 'Rosso', utilises multiple low-cost design innovations to obtain Doppler velocities for Sun-like and M-dwarf stars at <1 ms$^{-1}$ precision. The spectrograph has an asymmetric white-pupil format with a 100-mm beam diameter, delivering R>75,000 spectra over a 580–950 nm range for the Rosso channel. Simultaneous calibration is provided by a single-mode pulsed laser frequency comb in tandem with a traditional arc lamp. A bundle of 19 object fibres provides a 2.4" field of view for full sampling of stellar targets from the AAT site. Veloce is housed in dual environmental enclosures that maintain positive air pressure at a stability of ±0.3 mbar, with a thermal stability of ±0.01 K on the optical bench. We present a technical overview and early performance data from Australia's next major spectroscopic machine.

## 1. INTRODUCTION

Planet detection via high-resolution Doppler spectroscopy of stars is becoming increasingly popular in the astronomical community. While high-resolution spectrographs are nothing new, it is the intrinsic stability of these instruments that has improved in recent years, with instruments such as HARPS[1][2], and soon ESPRESSO[3], all but eliminating instrument variations over very long timescales.

Veloce is a stabilised fibre-fed echelle spectrograph for the 3.9 m Anglo-Australian Telescope (AAT) at Siding Spring Observatory. In a departure from other precision radial velocity instruments, Veloce is based on the philosophy of "just-enough-stabilisation". Rather than design a system that never changes, the aim is to achieve stability levels that result in changes sufficiently small to be reliably treated as differential and linear. These changes are then determined for every observation by the simultaneous measurement of both the science target and an ultra-stabilised calibration source, namely a laser frequency comb used in tandem with a traditional arc lamp. The primary driver for this approach is a limited budget, with ~75% of Veloce Rosso being funded by an Australian Research Council grant. The total project cost for the spectrograph, excluding the fibre cable and calibration laser, but including all engineering labour, was AUD ~1.65M.

Veloce 'Rosso' is the first of three spectrograph channels in the Veloce project and covers a wavelength range of 580–950 nm. The instrument relies on multiple design innovations to obtain Doppler velocities for Sun-like and M-dwarf stars at sub-m/s precision, but for a modest cost. Led by the University of New South Wales (UNSW) in Sydney, Veloce Rosso was designed and built at the Australian National University's Mount Stromlo Observatory facility near Canberra. The fibre feed system and instrument software are being provided by the Australian Astronomical Observatory (AAO) in Sydney. A system diagram showing the various instrument sub-systems is shown in Figure 1.

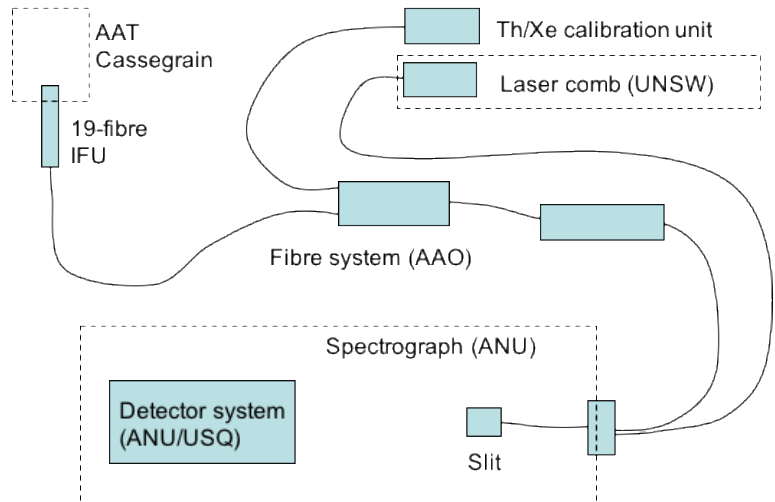

Figure 1: Instrument sub-system diagram showing institutional responsibilities: The spectrograph and detector system were built at the Australian National University (detector funding via University of Southern Queensland); the laser comb calibration unit is the responsibility of the University of New South Wales, Sydney; and the fibre system is by the Australian Astronomical Observatory.

## 1.1 Schedule and status

Veloce Rosso has been installed at the AAT and is currently awaiting integration of the fibre feed system. Figure 2 and Figure 3 show the completed installation. First light is scheduled for July 2018, including full commissioning on the 'VeloceCal' pulsed laser comb reference source. This paper is a technical description of the instrument, including early data on environmental stability and optical performance.

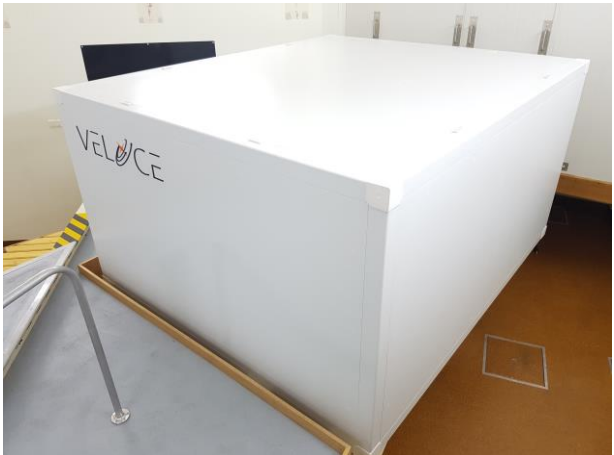

Figure 2: The Veloce instrument installed at the AAT, viewed from the north-west.

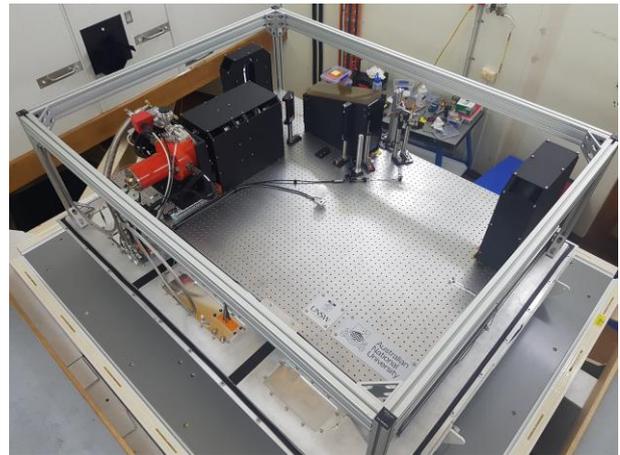

Figure 3: Veloce with external and internal enclosures removed, viewed from the south-west.

## 1.2 Radial velocity error budget

The radial velocity error budget for Veloce is summarised in Figure 4, arriving at a total single measurement accuracy of 0.5 ms$^{-1}$. This error budget is in a similar form to Ireland et al. (2016)[4], which was in turn inspired by Podgorski et al. (2014)[5], but where there is a more conventional approach to relying on fibre scrambling rather than simultaneous slit profile imaging.

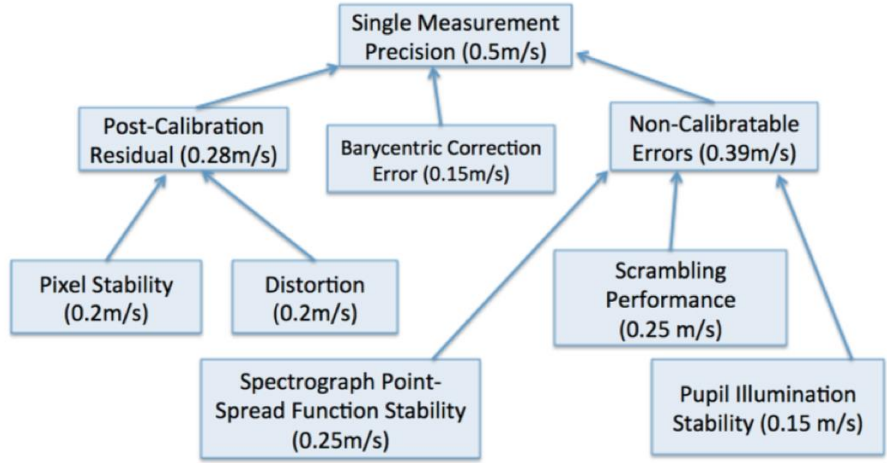

Figure 4: Radial velocity error budget for Veloce Rosso.

The Veloce error budget relies on a simultaneous reference source, and does not require highly sub-pixel wavelength scale stability. The error terms most relevant to the spectrograph are the Spectrograph Point Spread Function (PSF) Stability, and the influence of the PSF on the Pupil Illumination Stability. As described in Ireland et al. (2016)[4], the difference in line spread functions between stellar lines and reference lines means that subtle changes in PSF shape shift the measured stellar cross-correlation with respect to the reference source. Most importantly, this results in a requirement for a focal plane stability of better than 15 µm. This is in turn guaranteed by a ±0.2 K temperature stability of the entire spectrograph. An allowance for worse-than-expected PSF components (especially coma) and the influence of distortion on extracted radial velocities results in an overall requirement of ±0.1 K for the temperature stability of the entire spectrograph.

## 1.3 Technical requirements

The top-level measurement goal for Veloce Rosso is for single measurement accuracies of <0.5 ms$^{-1}$. This flows down to the set of technical requirements summarised in Table 1. Besides wavelength range, resolution and throughput, the principal engineering goals concern keeping systematic errors low. Two terms dominate here: i) temperature stability, to control spectral drift and defocus due to thermal expansion; and ii) pressure stability at the echelle grating surface, to provide constant refractive properties.

At this stage in the project (between installation and commissioning), performance measurements are not final. This paper will present early results for various subsystems, as per the numbers in the right-most column of Table 1.

Table 1: Summary of the major technical requirements for Veloce Rosso, with early estimates of achieved performance in the right-most column.

| Parameter | Requirement | Current performance |
|---|---|---|
| Wavelength | At least 588–900 nm | 580–950 nm |
| Resolution | At least 75,000 | 63,000–86,000 (TBC) See section 9.1 |
| Spectrograph throughput (with fibre) | At least 12.7% at order centres in the range 600–800 nm, and at least 10% at order centres in the range 588–900 nm. | 20–40% (without fibre) See section 9.2 |
| Fibre bundle throughput | At least 63% at order centres in the range 600–800 nm, and at least 60% at order centres in the range 588–900 nm. | TBD |
| Temperature of entire spectrograph | Constant to within ±0.1 K (goal ±0.01 K) of a set point. | ±0.05 K See section 6.2 |
| Air pressure at echelle grating surface | Stable to within ±5 mbar (goal ±1 mbar) of a set point. | ±0.3 mbar See section 6.1 |
| Expandability | The spectrograph will be built for an additional two cross-dispersed arms. | In-built space for green and blue channels |

# 2. OPTICS

## 2.1 Optical design

The Veloce spectrograph has been designed to deliver a resolving power of $R \geq 75{,}000$ for a 107 µm wide (90 µm FWHM) projected slit with 19 object and 9 calibration fibres. The optical layout is shown in Figure 5.

The echelle disperser is a Richardson R4 echelle grating, from master MR263, with a design blaze of 76º, a line frequency of 31.6 ln/mm, and coated with protected silver. The 100-mm pupil is compressed by a factor of 2.4 by the pupil transfer mirror. Cross-dispersion is performed with a 835.344 ln/mm slanted fringe VPH grating used off-Littrow. This creates a spatial:spectral anamorphic pupil scaling of 1:1.74. The result is a ~2-px FWHM in the spectral direction and ~1.15-px FWHM in the spatial direction. The refractive f/2.4 'Rosso' camera uses a horizontally-decentred lens element to control the astigmatism in the system. A 4k×4k e2v CCD with 15 µm pixels is used at the image plane to deliver an increased spectral grasp beyond the free spectral range of the echelle, up to 125% at order 103. The detector system is discussed in more detail in section 3.

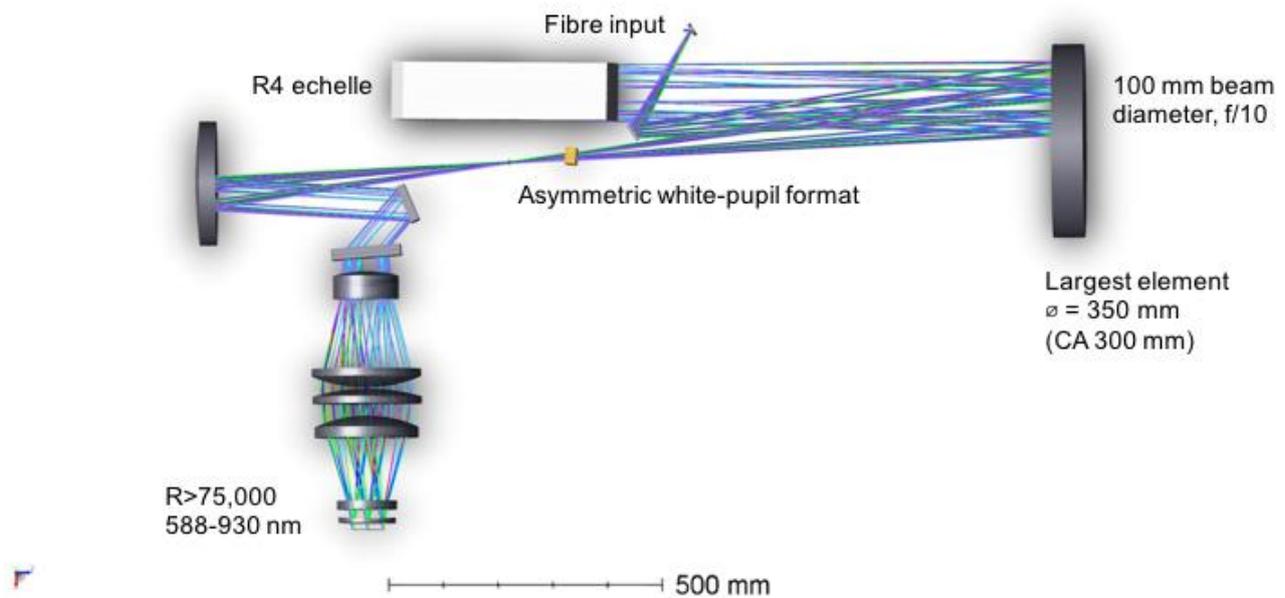

Figure 5: Spectrograph layout showing a ray-trace of the Rosso channel (580–950 nm) without the fibre relay optics

## 2.2 Opto-mechanics

Veloce uses in-house manufactured aluminium mounts for almost all its optics. This decision was made to avoid commercial off-the-shelf offerings with too many adjustable axes, to restrict unwanted motion as much as possible. The collimator mirror and pupil-transfer mirror both rest on horizontal posts and are held in place via Bellville washer stacks.

The Rosso channel camera assembly uses a somewhat non-standard approach in that individual lenses are held – again on horizontal posts – within a 'cage' assembly. This exterior of the assembly can be seen later in Figure 6. The design philosophy here was to precisely measure the (pinned) assembly before manufacturing the lens mounting posts, so that manufacturing errors could be corrected and all lens elements became aligned by tolerance. The design was also intended to present a cost-effective camera system. In practice, issues were encountered that are the subject of a paper by Herrald et al. (2018)[6]. Individual lens cells were manufactured to reduce mounting stresses, in turn forcing remanufacture of the mounting posts.

## 3. DETECTOR SYSTEM

The Veloce Rosso detector system inherits many features from the WiFeS spectrograph on the ANU 2.3 m telescope at Siding Spring Observatory[7]. The re-use of designs from WiFeS was chosen over a commercial off-the-shelf detector solution due to the risk of variable heat loading from a packaged unit with built-in electronics.

### 3.1 Detector and cryostat

The detector cryostat is shown installed in Figure 6 and open in Figure 7. The detector is an e2v CCD231-84-1-E74 4k×4k device with 15 µm pixels and e2v's 'astro multi-2' coating. The detector is mounted on a flat plate that is supported by a composite truss with minimal thermal variability. The detector mounting plate is thermally coupled to a cold head at the rear of the cryostat via a copper braid. This braid provides in-built protection, in that the detector cannot cool down or warm up too quickly via this thermal path. The cold head is part of a Polycold PCC Compact Cooler system from Brooks Automation with the compressor sited in another room at the end of ~50 ft lines. A coconut charcoal getter is directly attached to the cold head. The temperature of this assembly reaches around 85 K during operation.

Veloce's reliance on a stable environment means that there must be minimal disturbance to the instrument when pumping down a detector cryostat. There must also be negligible heat introduced into the environment. As such, the detector system has a remote vacuum valve that is pneumatically operated, with fore line bellows and a pilot air line exiting the instrument through hermetic feedthroughs. A modest MKS MicroPirani+piezo vacuum gauge is attached to the cryostat, to allow on-demand (i.e. infrequent) monitoring of pressure levels with minimal power dissipation.

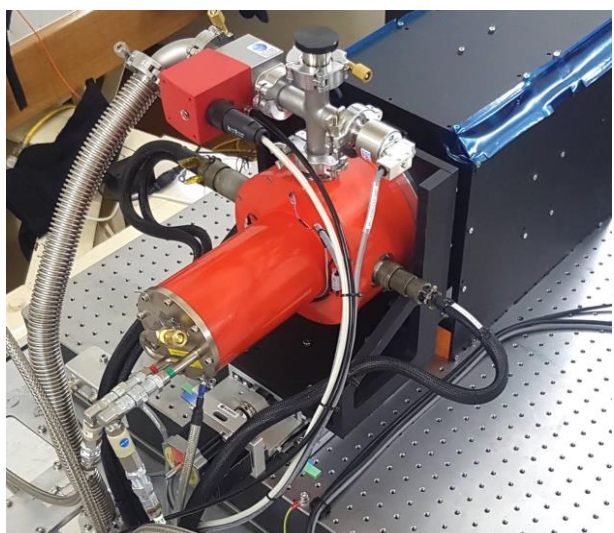 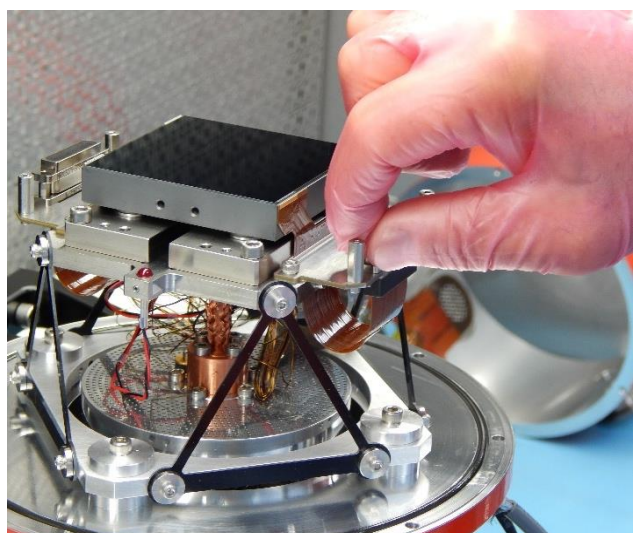

Figure 6: The installed Veloce Rosso cryostat, with remote vacuum valve (red box), pressure relief valve (top), and compact vacuum gauge (top-right). Refrigeration lines enter at the back.

Figure 7: The interior of the cryostat with its large outer ring removed, showing the charcoal getter (silver disc), copper braid (centre), black composite truss, and detector assembly (top).

### 3.2 Control electronics

Detector control is via an SDSU Gen III controller from Astronomical Research Cameras Inc. (ARC). This solution was chosen due to heritage and experience with past systems, including the ANU's WiFeS spectrograph. The controller is housed in a separate climate-controlled enclosure, directly underneath Veloce. This keeps cable lengths short while reducing heat load on the room.

Detector temperature regulation is handled by a Lake Shore Model 336 controller driving heaters on the detector mounting plate. Feedback is provided by Lake Shore calibrated temperature sensors. The temperature setpoint for the detector assembly is 145 K, and has a typical stability of ±0.03 K. This is well within the required stability of ±1 K, derived from the influence of thermal expansion on pixel size and position with respect to Veloce's radial velocity error budget.

Veloce has multiple detector readout modes, which are summarised in Table 2 along with typical characteristics. These modes are still being finalised at the time of writing.

Table 2: Preliminary Veloce Rosso detector readout modes, including a high dynamic range (HDR) mode for bright targets.

| Mode | Pixel rate (kHz) | Gain (ADU/e$^-$) | Full well (e$^-$) | Read noise (e$^-$) |
|---|---|---|---|---|
| Slow | 100 | ~1.0 | ~65,000 | <4.0 |
| Faster | 344 | ~1.5 | ~98,000 | TBD |
| Fastest | 709 | ~2.0 | ~130,000 | TBD |
| HDR | 100 | TBD | ≥150,000 | TBD |

## 3.3 Cryogenic detector preamps

The ANU has been actively developing cryogenic preamp stages for infrared detectors and, more recently, CCDs. A cryogenic preamp stage offers superior noise immunity by placing an initial gain stage very close to the detector, enabling longer cable runs and/or reduced pickup noise from the surrounding environment. Cryogenic preamps for the Veloce Rosso detector system were not included in the final instrument due to scheduling limitations, but may be added in future. More information can be found in Vaccarella et al. (2018)[8].

## 4. FIBRE FEED

The Veloce fibre feed is the subject of another paper by Case et al. (2018)[9]. The fibre feed system consists of three main subsystems: i) the fore-optics, which includes the optical barrel and the microlens array (MLA); ii) the fibre cable, which includes both the integral field unit (IFU) and slit fibre arrays, the fibre agitator, the spectrograph feedthrough and the splice box; and iii) the slit relay, which includes the slit relay optical barrel and the microlens array design. The IFU and slit format are shown in Figure 8.

The Veloce spectrograph concept relies on efficient slicing of the input feed into three sections in the dispersion direction. This is achieved at the telescope focus through image slicing with the MLA. The array feeds a bundle of 19 object fibres in a hexagonal pattern at the input, which is fanned out to a row of fibres at the spectrograph end. Additionally, five sky fibres and two calibration fibres are used, plus two dummy fibres, bringing the total number at the slit to 28. The MLA uses a standard pitch of 250 µm, resulting in a total input slit length of 7 mm at f/20.9. This reduces to 3.35 mm at f/10, the acceptance f-ratio of the spectrograph collimator.

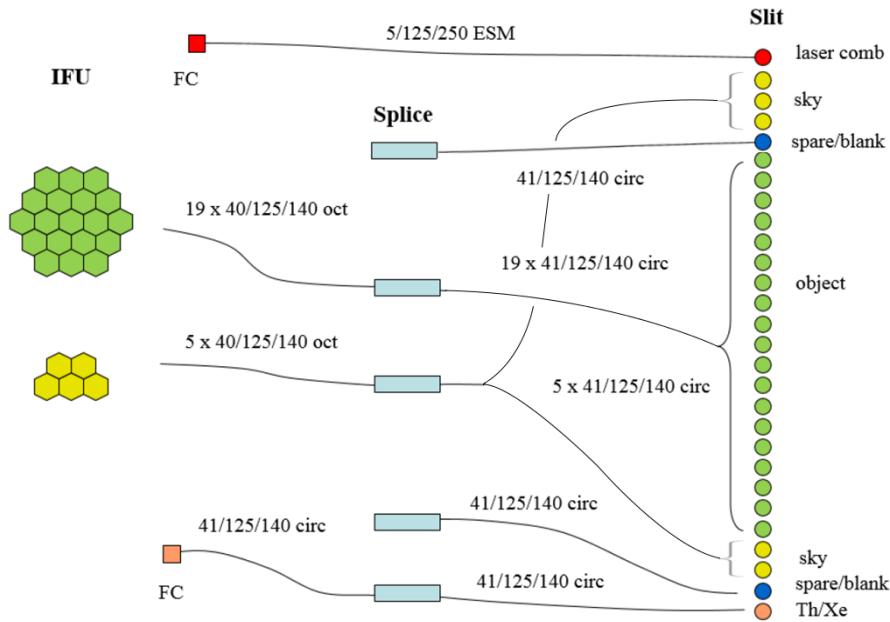

Figure 8: Diagram showing Veloce's IFU format and slit mapping. There are 19 object fibres (field of view 2.4") for full sampling of stellar PSFs in imperfect seeing conditions, five sky fibres, and two continuous calibration sources.

## 4.1 Fore-optics

The fore-optics system comprises a set of hexagonal MLAs, the first as a field lens and a relay preceding it that matches the telescope's Cassegrain focal ratio of f/7.9 to the desired f/25.5 for the fibre size and f-ratio. The fibres are fed at f/4. The fore-optics mechanical design, shown in Figure 9, is straight-forward. It features a lens barrel with a rear connector on a flange. The rear connector kinematically mounts the IFU head to the barrel. The lenses are first centred into cells, and the cells are then centred into the lens barrel. The system is then mounted onto CURE, the AAT's Cassegrain interface, aligned, reassembled, and can from that point be easily removed and re-attached to the telescope.

The mechanical design of the IFU, shown in Figure 10, follows the IFU design for the GHOST high resolution spectrograph on Gemini[10]. All parts are exact replicas except for the IFU body and the MLAs. The latter parts are substituted with the Veloce MLAs. There is also a change in the spacing between the two arrays, and a new fibre array substrate design.

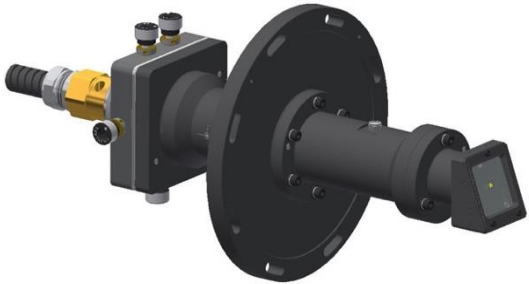
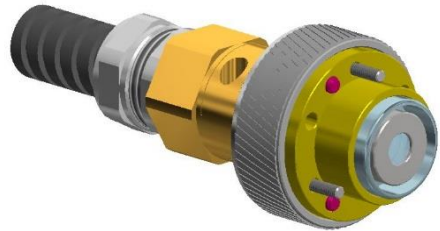

Figure 9: The Veloce fibre feed fore-optics assembly. This assembly attaches to the AAT's Cassegrain focus via a standard instrument interface.

Figure 10: The IFU heads on the Veloce fibre cable are based on those used for the GHOST spectrograph on Gemini.

## 4.2 Fibre cable

The IFU head attaches to the fore-optics relay. The IFU consists of the fibre array assembly (precision mounted fibres) bonded to the air spaced MLA pair. The fibre array is manufactured using femtosecond laser exposure and wet-etching of fused silica substrates, allowing for rapid manufacture at much lower cost than traditional photolithographic etch methods in silicon. The slicers themselves are based on two telecentric MLAs in series; one strongly curved lens set that focusses the beam, and a field lens set that achieves telecentricity.

Octagonal fibres have been chosen for the telescope injection side of the cable, to provide enhanced scrambling properties. The fibre size of 40 µm (flat-to-flat) was derived from analysis. Circular fibres are used for the spectrograph injection side of the optical cable, with core size of 41 µm. The chosen vendor is Polymicro Techologies (part of Molex corporation). This vendor has consistently delivered low-scattering optical fibre to specification on many previous AAO projects. The cladding diameters on both the octagonal and circular fibres is 125 µm and the polyimide coating outer diameter is 140 µm.

The agitator for Veloce will re-use many of the parts from the AAT's CYCLOPS2 instrument[11] with a large, low frequency oscillation and a small, high frequency oscillation to provide mode scrambling in the octagonal core fibre. A splice box is used to join the octagonal core fibre in the IFU section of the cable to the circular core fibre going to the spectrograph slit. These fibres are spliced using standard arc fusion splicing techniques.

The slit is a linear version of the fore-optics unit featuring a linear array of fibres bonded to an air-spaced pair of MLAs. The slit assembly will look the same as the IFU assembly. It will use all parts as for the GHOST version except for the IFU body (now a slit body), the MLAs, and the spacing between the MLAs. The slit body will be fabricated using the same approach as the IFU body.

### 4.3 Slit relay

The output configuration of the fibre cable for injection into the spectrograph is similar to the input, but in reversed order. However, the MLA is a linear array, demanding a large field of view in the relay optics that follow (7 mm at f/20.9). To account for some focal ratio degradation and to ensure that edge effects on the MLA are kept at bay, we use a fibre image size on the 'slicing' MLA of 228 µm.

For the input slit relay, two doublet-doublet configurations and a doublet-triplet configuration were investigated, all of which are shown in Figure 11. The modified doublet solution (the middle option) was found to be sufficient in terms of image quality and beam movement and thus was baselined for this design.

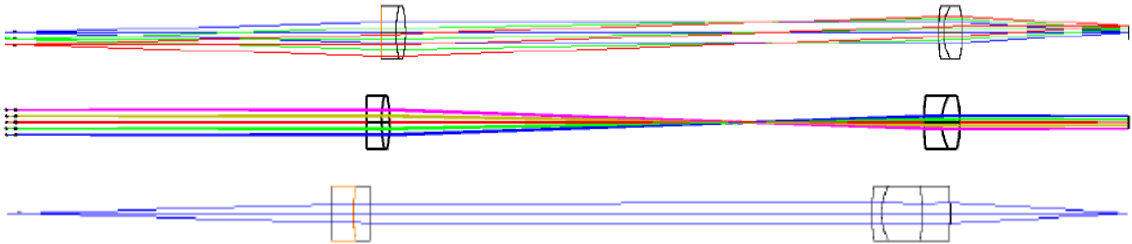

Figure 11: Three optical relay designs were considered for the Veloce input slit. The chosen baseline design was the second dual doublet configuration (middle drawing).

## 5. ENVIRONMENTAL ENCLOSURES

Veloce's optical bench is housed inside two enclosures, simply named 'internal' and 'external'. The purpose of these enclosures, shown in Figure 12 and Figure 13, is to provide a light-tight environment and to facilitate temperature and pressure stability. Temperature and pressure are actively regulated via dedicated heaters, an air conditioning system, and a pressure control system.

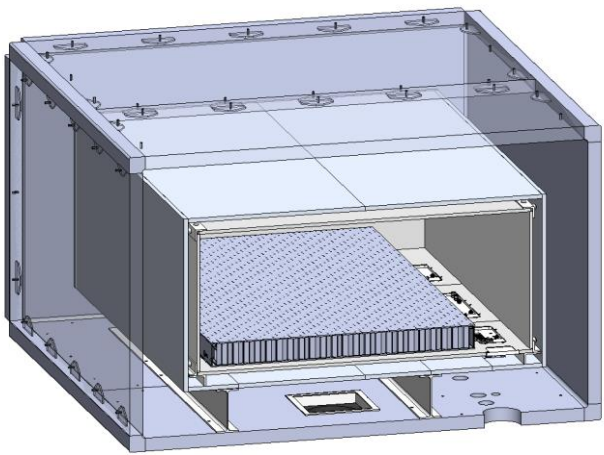

Figure 12: A CAD model section view of the Veloce optical bench inside the 'internal' and 'external' enclosures.

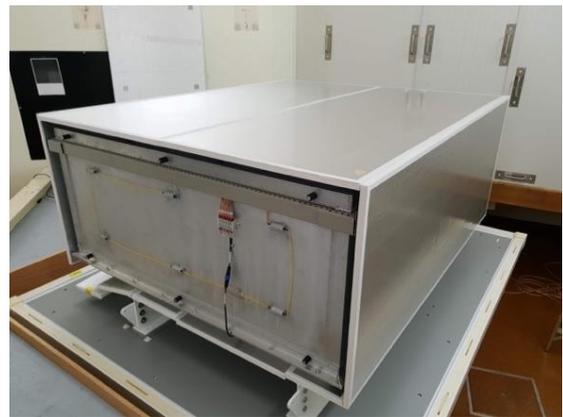

Figure 13: A foam panel has been removed to reveal a heater circuit on one of the internal enclosure panels. The external enclosures walls have been removed.

## 5.1 Internal enclosure

The internal enclosure, as the name implies, is the first layer around the spectrograph. It is effectively an air-tight box split around the perimeter of the base to allow removal and access. The upper lid is secured to the base by means of M8 bolts and nuts that press the flanges onto a flat rubber seal.

The design of this enclosure was driven by Veloce's pressure stability requirement, which does not prescribe whether the absolute pressure of the instrument should be positive or negative. While a vacuum system presents technical benefits in terms of thermal control, this was not considered feasible due to budget constraints. Instead, the internal enclosure maintains a positive pressure slightly above that of the ambient environment. The enclosure was designed to operate at a differential pressure of up to 30 mbar (0.435 psig), based on historic records of barometric pressure at the observatory (minimum 879 mbar; maximum 901 mbar) and a setpoint just above the maximum (905 mbar, see section 6.1), plus a margin. The actual failure point of the enclosure is significantly higher.

The enclosure is made from large aluminium honeycomb panels bonded inside a welded aluminium frame and sealed all around. The larger top and bottom surfaces must withstand a maximum equivalent load of ~1 tonne. Indeed, during operation it is possible to witness the top panel bulge by up to 5 mm at the centre. The optical bench kinematically rests on the lower panel by means of three contact pads, for consistent support.

Air-tight cable penetrations into the enclosure are achieved via elastomeric feedthrough panels from Roxtec. This option was selected as a way of minimizing breaks in the noise-sensitive detector cables, and to provide flexibility in future cable allocation. While an adequate seal can be achieved by this method, in practice this approach has been problematic due to limited working space and the need to rebuild the entire assembly if one cable changes. Vacuum, air and refrigeration lines use dedicated bulkhead connectors. The Rosso channel feedthrough panel is shown in Figure 14.

The term 'air-tight' is used rather loosely in the context of this instrument, as there are known leaks in the enclosure. These are primarily around the Roxtec feedthroughs. The total measured leak rate is ~1.5 L/min, which is considered tolerable.

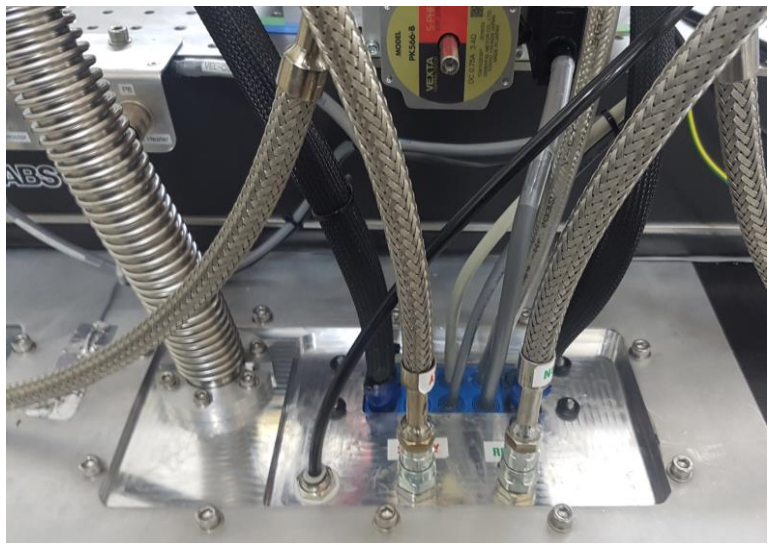

Figure 14: An example of one of Veloce's sealed internal enclosure feedthrough panels. The Roxtec cable feedthrough system is the blue block at the rear of the panel. A vacuum hose, refrigerant hoses and compressed air penetrate the panel via dedicated bulkhead connectors.

Electrical heaters are fixed to the outer aluminium shell using thermal epoxy as part of the instrument's thermal control system (see section 6.2). These are simply power resistors arranged in multiple series circuits around the main panels, as seen in Figure 13. A layer of 19 mm thick polyisocyanurate (PIR) thermal insulation tightly wraps around the internal enclosure in the form of rigid foam panels. Small metal plates are glued to the rear of the panels, allowing them to be held in place by magnets on the internal enclosure.

## 5.2 External enclosure

The external enclosure acts as the main outer layer of insulation required to run the temperature control system. This enclosure is air conditioned to a temperature slightly lower than the instrument temperature (see section 6.2), via ducts in the enclosure floor. The enclosure is constructed from commercially available cold room panels made by MISA. The outer dimensions are 2550×2150×1350 mm (L×W×H) and a ~200 mm air gap is maintained between both enclosures. All panels have a 60-mm thick foam core sandwiched between painted steel sheets and joined together with a cam-lock system. The floor panel of the external enclosure can be seen in Figure 13. The entire instrument is bolted onto a large vibration-isolated steel platform inside the Coude 1 room of the AAT. Figure 2 on page 2 shows the complete assembly.

## 5.3 Air conditioning system

The external enclosure is temperature-regulated by a custom air conditioning system based on a Huber Minichiller 600-H thermal control unit. The system's water circuit is split into two to allow the Huber unit to reside on the service floor of the telescope and away from the instrument. The Minichiller's low pressure circuit runs only a short loop, feeding into a small Lytron liquid to liquid heat exchanger (LL510G12). The recirculating pump high pressure circuit runs a longer loop from the service floor to a Lytron liquid to air heat exchanger (6220G1SB) near Veloce. This heat exchanger is coupled to a fixed-speed fan that circulates the external enclosure air via a pair of insulated flexible ducts.

# 6. ENVIRONMENTAL CONTROL

Two aspects of environmental control are crucial for Veloce: i) pressure control, to maintain a constant refractive index at the surface of dispersive optics; and ii) temperature control, to maintain the position of opto-mechanical systems in terms of thermal expansion. These two factors are controlled by separate systems that are summarised in this section. More details can be found in Gausachs et al. (2018)[12].

## 6.1 Pressure control system

The absolute pressure in Veloce's internal enclosure must be regulated to maintain the refractive behaviour of the instrument's echelle grating. The pressure stability requirement was defined as ±5 mbar, with a goal of ±1 mbar.

Veloce operates at a positive pressure with respect to its ambient environment (see section 5.1) and the chosen pressure setpoint for the altitude of the installation site is 905 mbar. Dry compressed air is used to 'pump up' and maintain the instrument pressure. The air supply is controlled by an industrial pressure controller from Alicat Scientific. This controller has shown excellent performance over many months of laboratory and integrated testing, comfortably maintaining ±0.3 mbar stability in Veloce's 1800 L internal enclosure despite rapid changes in ambient barometric pressure and temperature. The controller is part of Alicat's PCD range, but with a custom modification to allow more accurate pressure feedback via a remote barometric sensor. Figure 15 shows demonstrated pressure performance over 600 h.

## 6.2 Thermal control system

The driving technical requirement for the thermal control system is a temperature stability of ±0.1 K on the optical bench, with a goal of ±0.01 K. As described in section 5, this is achieved via nested enclosures that decouple the spectrograph from seasonal temperature variations in the room with a combination of insulation and active temperature control.

Veloce's external enclosure has a ducted air conditioning system that regulates the air temperature to ~24 °C through heating and cooling. A sensor inside the enclosure acts as the feedback for a control loop that actively adjusts the internal setpoint of the Huber unit as necessary. This accounts for any losses in the heat-exchange circuit and has been shown to maintain an air temperature stability of ±0.1 °C at the time of installation, as shown in Figure 16.

Veloce's internal enclosure features several resistive heater circuits on its exterior. These heaters act to hold the temperature of the optical bench inside at a temperature of 25 °C, i.e. higher than that of the air in the external enclosure. These heater circuits are powered by a 24 V DC supply, using pulse-width modulation (PWM) to control their power. The entire heating system has a maximum power of 96 W, though only a fraction of this is used to maintain the temperature setpoint.

The heaters are split into five separate circuits based on location: i) the enclosure base; ii) the enclosure ceiling; iii) the long walls of the enclosure; iv) the short walls of the enclosure; and v) on the detector cryostat body. Feedback is via calibrated thermistor pairs on the base, ceiling, cryostat, and the optical table (where the master setpoint applies). These separate loops operate as a cascaded proportional-integral-derivative (PID) control system to ultimately maintain the optical table temperature at 25±0.1 °C or better. Figure 16 shows demonstrated stability well within the goal of ±0.01 K.

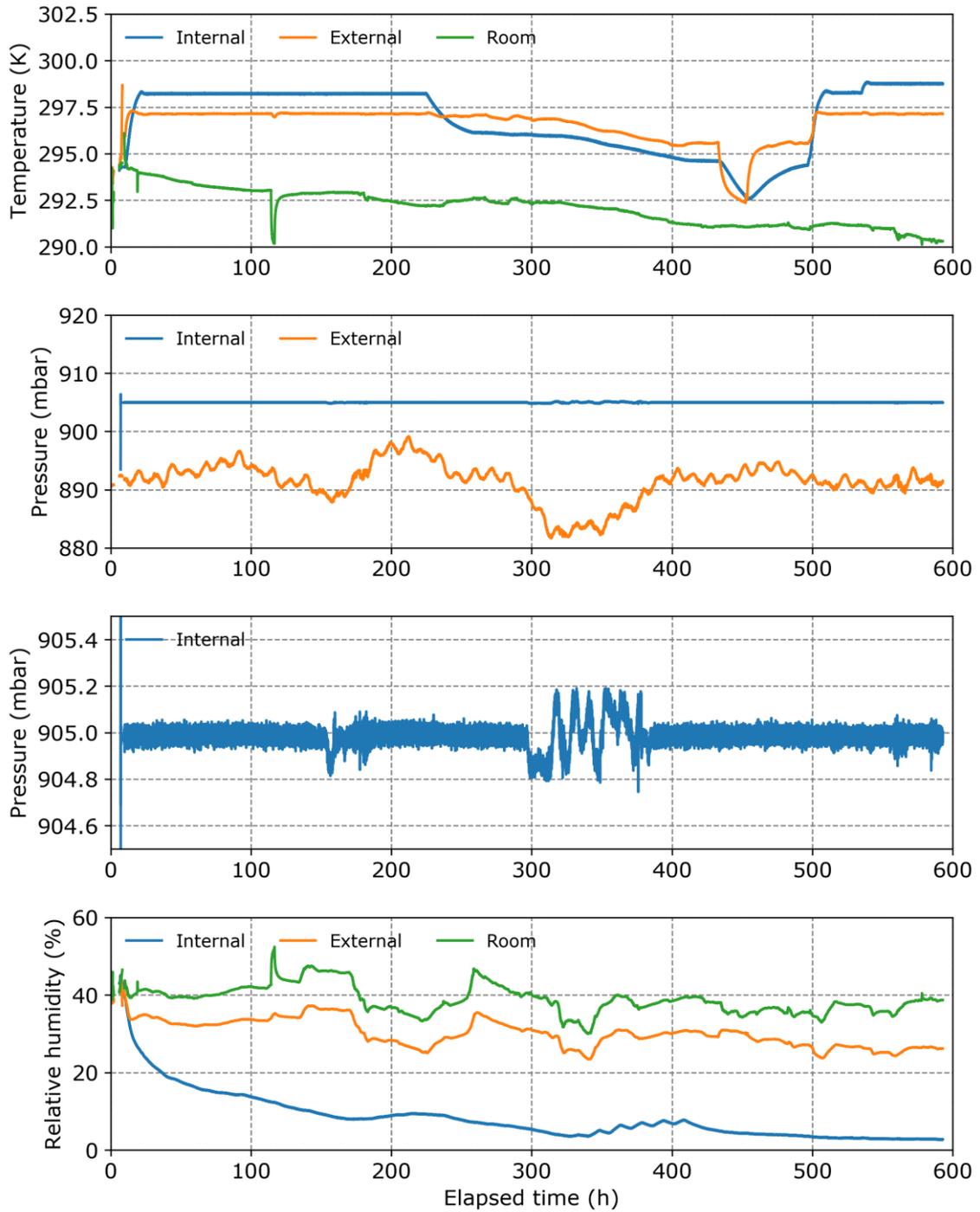

Figure 15: Pressure control testing over ~600 h. The third panel shows an internal enclosure pressure stability comfortably within ±0.3 mbar of the 905 mbar setpoint (requirement 5 mbar, goal 1 bar). This is despite swings in ambient temperature (first panel) and pressure (second panel) throughout the test. Note the decline in internal enclosure humidity as Veloce's internal enclosure air is gradually replaced by the dry air supply (due to small leaks).

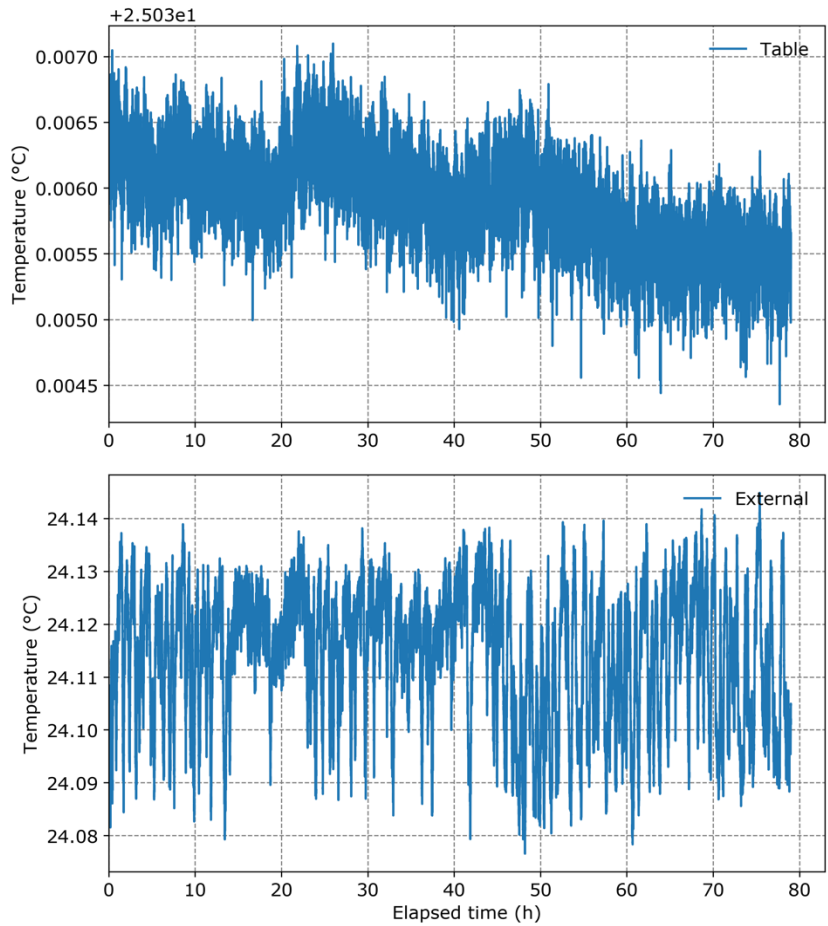

Figure 16: Veloce's thermal control system can maintain the temperature of the spectrograph's optical table within the ±0.01 °C stability goal. The external enclosure air conditioning system (bottom panel) is set ~1 °C cooler and has a stability of ±0.1 °C. The small offset from each of the setpoints have since been removed by calibration.

The thermal control system's digital interface is a LabJack T7 Pro laboratory controller. This was chosen for its low cost and abundance of I/O channels, including high-resolution (24-bit) differential analogue-to-digital converters. The Veloce control computer communicates with the LabJack interface over Ethernet, allowing remote reading of all thermistors and PWM operation of the heater circuits (with added MOSFET drivers).

## 7. SOFTWARE

The Veloce control system software has a client-server architecture and is built on object-oriented design principles. Each server supports one or more pieces of hardware, and clients communicate with the server via structured request-reply messages using 0MQ middle-ware. The servers use polymorphism to specialise interfacing with specific hardware, such as the ARC detector controller, while providing a common interface for the client-server communication framework.

Three separate servers support the core functions of the instrument. One for the detector controller, one for thermal control, and a third for the focus mechanism and sensor monitoring.

The server supporting the focus mechanism and sensors conglomerates a bespoke array of Ethernet enabled devices. The core processing thread obtains data from the devices, using ModbusTCP where supported, and updates the server's process image. Multiple concurrent client requests are serviced from the process image in a thread-safe manner.

The Veloce control computer presents a command line interface for operating and monitoring the instrument during engineering operations such as installation and initial commissioning. This will be replaced by a graphical front-end, currently being developed by the AAO.

# 8. CALIBRATION

Veloce Rosso's design is based on a philosophy of "just-enough-stabilisation", as opposed to hyper-stabilisation. Rather than design a system that *never* changes, the aim is to achieve pressure and temperature stability levels that result in changes sufficiently small to be reliably treated as differential and linear. These linear differential changes are then determined *for every observation* by the simultaneous observation of both science targets and an ultra-stabilised calibration source.

The principal calibration source chosen for Veloce Rosso is an FC1000 femtosecond pulsed-laser Astrocomb from Menlo Systems GmbH. Originally planned as an upgrade to Veloce, the system has the useful property of being available before the instrument is commissioned, allowing the use of all reference sources at first light. The laser comb will be used in tandem with a traditional Thorium-Xenon absolute reference source.

The Astrocomb system is driven by a laser pulsed at a timing accuracy of better than 1-part-in-$10^{11}$ (equivalent to a velocity precision of 3 mms$^{-1}$), when slaved to a GPS-mediated quartz clock. This delivers a mode spacing of 250 MHz, which is mixed down to optical wavelengths, and broadened to cover the range 450–950 nm at an output power of >100 μW. Three laser-locked cavities are then used to remove unwanted modes, delivering a comb with a mode spacing of 25 GHz, suitable for use with an *R*=80,000 echelle spectrograph.

Output from the laser comb is delivered by an endlessly single-mode photonic crystal fibre. In a departure from previous practice, Veloce Rosso will be fed *directly* with this single mode fibre, rather than attempting to deliver it via the same multi-mode fibres used for light from the telescope. The key insight here is that any calibration source (whether arc-line or laser comb) will always have a fundamentally different illumination from the science fibres. It is worth noting that these fibres only ever calibrate the stability of the spectrograph; the stability of the sky illumination relies on scrambling within the fibre feed and is fundamentally uncalibratable.

Veloce Rosso, therefore, will use the (as far as the spectrograph is concerned) unresolved signal from the single-mode fibre to calibrate the spectrograph, *and* deliver simultaneous calibration of spectrograph optical performance and line-spread function with every science exposure.

# 9. SPECTROGRAPH PERFORMANCE

Veloce Rosso is yet to receive first starlight at the time of writing. This has limited testing to non-telescope sources such as laboratory arc lamps or the sun. Figure 17 shows an example of a solar spectrum raw frame, with a cursory reduction of a single order in Figure 18.

## 9.1 Resolution

Evaluation of optical resolution will be best achieved with the laser comb calibration source during commissioning (currently scheduled for July 2018) and by subsequent astronomical observations. Current resolution estimates have been derived from non-ideal spectral sources, but are presented here as an indication.

The resolution of the system was measured using a HgAr lamp fed into a single 105-μm diameter fibre. Pairs of inter order spectral lines were measured to determine resolution from known spectral lines. The FWHM was determined using a Gaussian curve fitting tool across a line profile, and hence the curve of the spectral tram lines was not considered. The measurement process is illustrated in Figure 19.

The equation used to measure the resolution from pairs of known spectral lines is

$$R = \frac{\lambda}{\Delta\lambda} = \frac{\lambda}{P \cdot dx_{FWHM}}$$

where *P* is the linear distance between the spectral lines and *dx* is the measured FWHM for an individual spectral line. In this case, the FWHM is determined from a line profile of the resolvable element fitted to a Gaussian profile.

The results shown in Table 3 give the measured Rosso channel resolution at four places in the echellogram, and range from *R*=63,117 to *R*=85,803. These measurements were limited by the spectral width of each line, which for HgAr is known to be non-negligible for this system. The background was not subtracted for these measurements and the FWHM was only fitted across a line profile of each spectrum. As such, we believe that the true resolution of Veloce Rosso will meet the *R*≥75,000 resolution goal across most of the image.

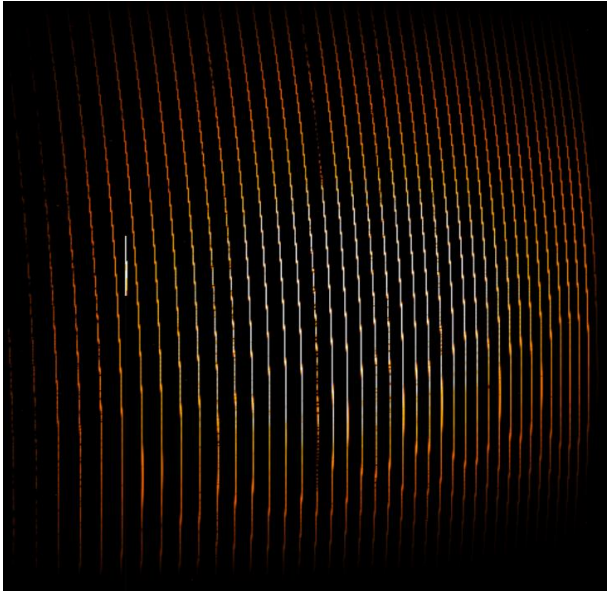

Figure 17: A raw Veloce Rosso solar exposure.

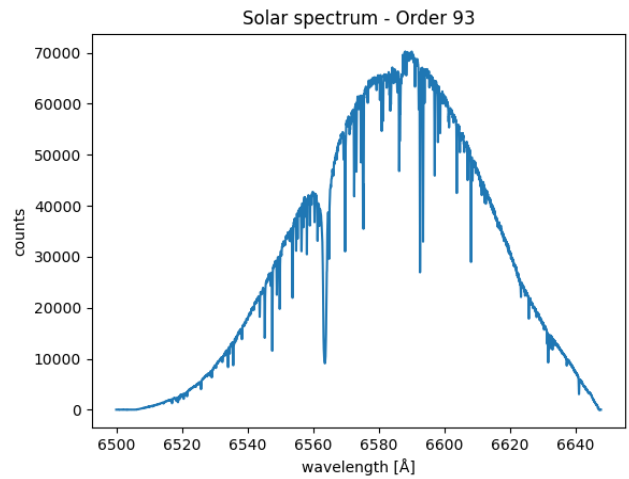

Figure 18: A 'quick look' reduction of one of the orders in Figure 17, showing clear absorption features.

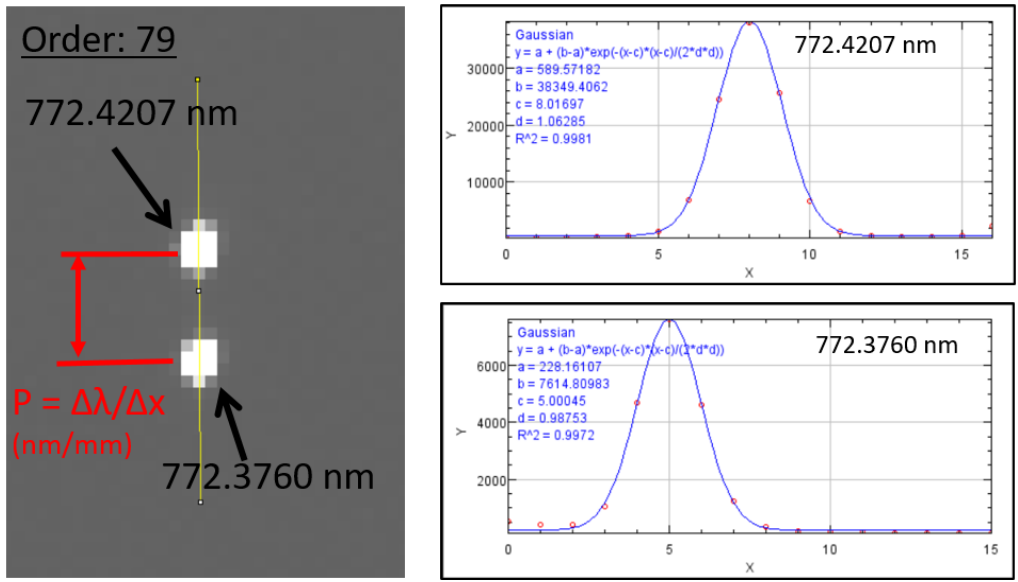

Figure 19: Spectral pairs at order 79 (772.4207 nm and 772.3760 nm), used to estimate the resolution of the system.

Table 3: Measured resolution of the Rosso channel for four HgAr spectral line pairs.

| image | Line | order | wave 1 | x (pix) | y (pix) | FWHM (pix) | wave 2 | x (pix) | y (pix) | FWHM (pix) | distance (mm) | P (Um/mm) | Spot size FWHM (pix) | FWHm (mm) | Resolution |
|---|---|---|---|---|---|---|---|---|---|---|---|---|---|---|---|
| 59 | Ar | 79 | 0.7724201 | 2075 | 2731 | 2.046 | 0.7723761 | 2075 | 2741 | 2.455 | 0.15 | 0.0002933 | 2.0459 | 0.030689 | 85806 |
| 59 | Ar | 81 | 0.7514652 | 2289 | 3121 | 2.352 | 0.7503869 | 2286 | 3334 | 2.342 | 3.20 | 0.0003375 | 2.342 | 0.03513 | 63297 |
| 59 | Ar | 73 | 0.8424647 | 1288 | 1312 | 2.957 | 0.8408209 | 1319 | 1702 | 4.12 | 5.87 | 0.0002801 | 2.957 | 0.044355 | 67808 |
| 59 | Ar | 75 | 0.8103686 | 1620 | 3347 | 2.463 | 0.8115311 | 1619 | 3134 | 2.245 | 3.20 | 0.0003638 | 2.245 | 0.033675 | 66234 |
| | | | | | | | | | | | | | | MEAN | 70786 |

Lamp: HgAr (Image 1, 100msec), pix = 0.015 mm

## 9.2 System efficiency

The spectrograph efficiency was measured using a single central fibre for the Rosso channel. The measured system excluded the fibre feed and final field flattener lens (mounted in the cryostat). A relative measurement was taken between the system output and a reference signal. The reference was a fibre-coupled laser-induced white light source from Energetiq (LDSL EQ-99FC) and was considered 'invariant' over the time taken to measure the test and reference signals. This allowed the reference signal and the system signal to be measured using compact off-the-shelf fibre spectrometers, fed by a two-inch integration sphere. This type of system has been shown to tolerate significant background noise, as the integration sphere minimises the impact of the stray light that usually has a strong directional component[13].

The measured system efficiency is shown in Figure 20 and exceeds the spectrograph throughput requirement at all wavelengths. This does not include the fibre feed, although when inferring the minimum throughput requirement for the spectrograph only, the fibre was assumed to have the minimum acceptable throughput. Since it is likely that the completed fibre feed will exceed minimum requirements by some margin, we anticipate that Veloce Rosso's end-to-end throughput will be significantly above the minimum requirement of 12.7% for the central 600–800 nm range.

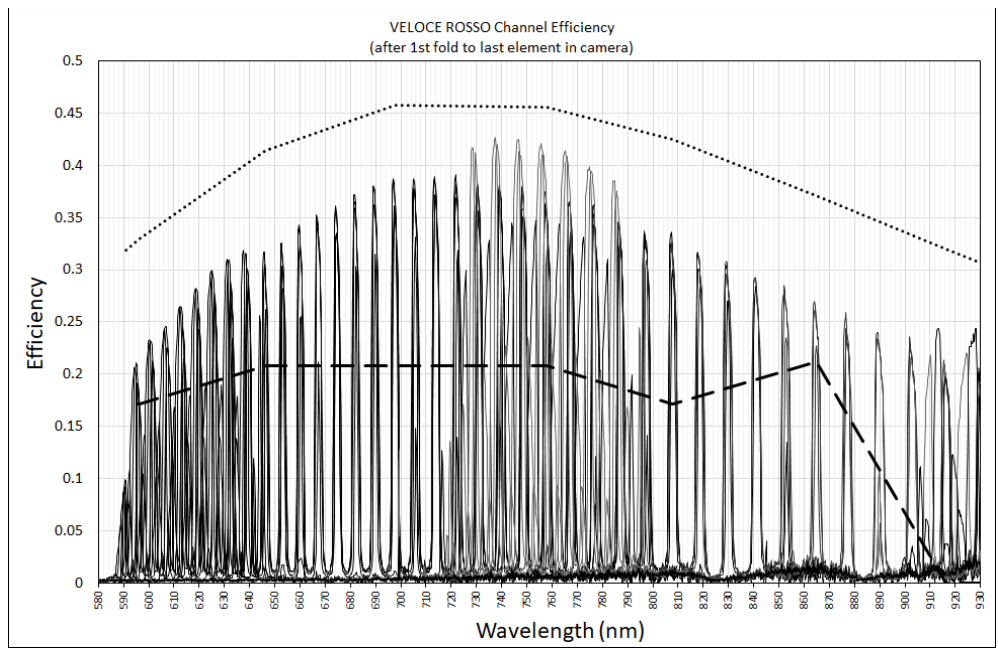

Figure 20: Veloce Rosso efficiency plot. The upper dotted line is the maximum expected design efficiency, the lower dashed line is the minimum required efficiency (assuming minimum fibre efficiency and theoretical detector quantum efficiency), and the solid line peaks show the as-built efficiency for the Rosso channel.

## 9.3 Slit losses and extraction

The input aperture is 2.4" in diameter, which provides the same etendue as the 1.2" diameter aperture for the GHOST spectrograph for Gemini South currently under construction[14]. The fiber feed borrows from the heritage of the CYCLOPS fiber optic cable and integral field unit for the UCLES spectrograph[11]. Veloce has additional sky fibers, so that it can be used to target faint, sky-limited targets as well as its primary bright exoplanetary science targets. Additionally, it has two simultaneous reference fibers (see Figure 8, p. 6), plus full coverage of the echelle free spectral range out to ~950 nm (in contrast to its predecessor UCLES, which only had full wavelength coverage below 530 nm).

Given the high etendue at the input of Veloce (equivalent to an 8-m telescope echelle spectrograph), the fiber profiles must be arranged on the detector with no space between them. Analysis techniques are being pursued both for extracting individual fibres (and accounting for cross-talk between them), as well as for extracting all fibres together (and modelling their fixed offsets and relative throughput to derive a cross-correlation template correct for every exposure). As a fall-back position, the fibres have been arranged on the pseudo-slit such that a mask over the outer hexagonal ring of 12 microlenses will allow the inner seven fibres to capture the core of the telescope image, and deliver a slit with at least two now-dark fibres in between illuminated fibres, allowing a traditional extraction of each illuminated core. This is illustrated in Figure 21 and Figure 22.

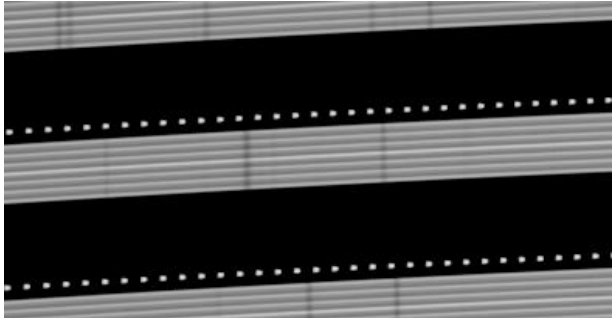

Figure 21: Simulated Veloce Rosso image with a full slit of 19 illuminated IFU fibres. The fibre core images overlap (darker cores are outer fibres).

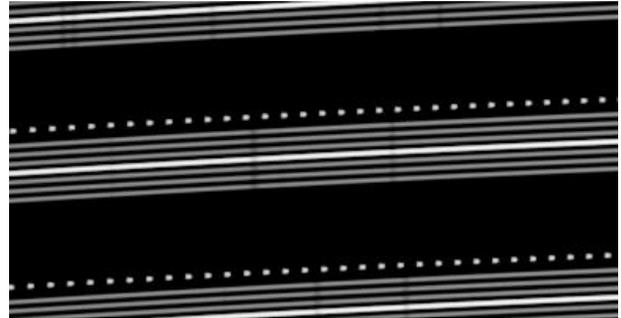

Figure 22: Simulated Veloce Rosso image with the outer ring of fibres covered at the input. The illuminated fibre images no longer overlap.

## 10. CONCLUSIONS & PROSPECTS

Veloce Rosso is a precision radial velocity spectrograph for the AAT that will be fully commissioned in the coming weeks. Early performance testing indicates good optical performance and stability. This instrument will deliver to Australian astronomers the capability to obtain high-precision Doppler velocities for the exoplanet host stars currently being discovered by southern hemisphere transit searches, and for the coming wave of discoveries by NASA's Transiting Exoplanet Survey Satellite (TESS).

Veloce's remaining two channels, 'Verde' and 'Azzurro', were successful in securing Australian Research Council funding for commencement in 2018. These additional arms will transform Veloce into a high-performance spectrograph covering the entire visible spectrum from 380 nm to 950 nm. The planned layout, which was accommodated for in the design of Veloce Rosso, is shown in Figure 23. This upgrade is expected to be installed by the end of 2019.

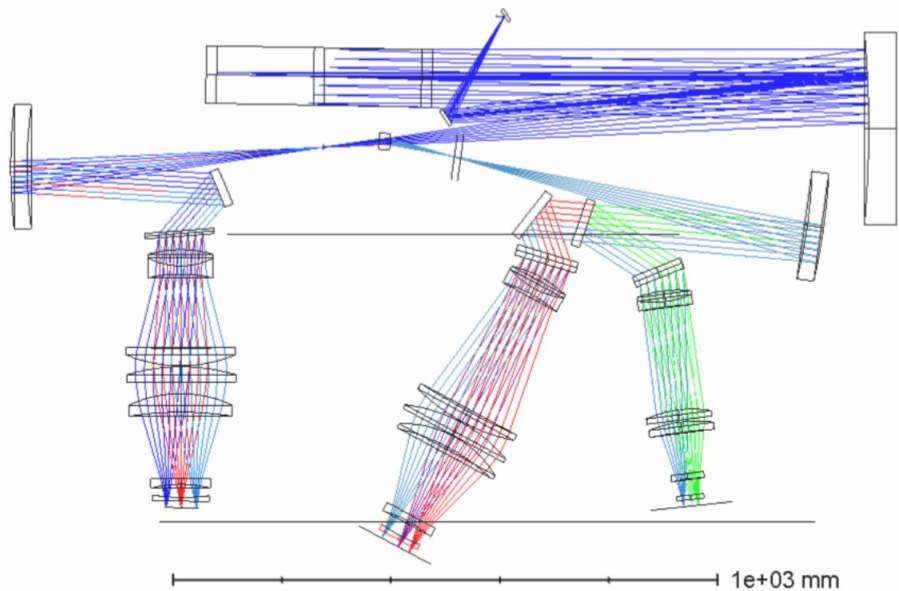

Figure 23: Two additional channels ('Verde' and 'Azzurro') will complete the Veloce spectrograph by the end of 2019.